\documentclass[runningheads]{llncs} 

\usepackage{amsmath} % text in equation 
\usepackage{amsfonts}  
\usepackage{setspace} % setstretch
\usepackage{graphicx} % resizebox
\usepackage{makecell} % for makecell
\usepackage{mathtools} % for \begin{multlined}
\usepackage{enumitem}
\usepackage[linesnumbered,ruled,noend]{algorithm2e}
\usepackage[misc]{ifsym}

\usepackage{caption} % for subfigure error
\usepackage{subcaption} % for subfigure error
\usepackage{mathtools} % for ceil
\usepackage{hyperref}
\usepackage{nameref}
\usepackage{xcolor}

\usepackage[maxnames=1, giveninits=true, doi=false, isbn=false, url=false]{biblatex}
\usepackage{tikz}
\usetikzlibrary{positioning}

\newcommand{\inull}[1]{}
\AtBeginDocument{%
  }


\begin{document}
\pagenumbering{gobble} % 排除页码
\newcommand{\iG}{\mathcal{G}}
\newcommand{\iV}{\mathcal{V}}
\newcommand{\iE}{\mathcal{E}}

\title{\texttt{ShareDP}: Finding k Disjoint Paths for Multiple Vertex Pairs}

% example
% \author{First Author\inst{1} \and
% Second Author\inst{2,3} \and
% Third Author\inst{3}}
% \authorrunning{F. Author et al.}
% \institute{
% xxx\\
% \email{lncs@springer.com}\and
% xxx\\
% \email{lncs@springer.com}\and
% xxx\\
% \email{lncs@springer.com}
% }             

\author{Zhiqiu Yuan\inst{1} \and
Youhuan Li\inst{2} \and
Lei Zou \Letter\inst{1} \and
Linglin Yang\inst{1}}
\authorrunning{Z. Yuan et al.}
\institute{
Peking University\\
\email{\{yuanzhiqiu, linglinyang\}@stu.pku.edu.cn, zoulei@pku.edu.cn} \and
Hunan University\\
\email{liyouhuan@hnu.edu.cn}
}
% \institute{
% Peking University\\
% \email{yuanzhiqiu@stu.pku.edu.cn} \and
% Hunan University\\
% \email{liyouhuan@hnu.edu.cn} \and
% Peking University\\
% \email{zoulei@pku.edu.cn} \and
% Peking University\\
% \email{linglinyang@stu.pku.edu.cn}
% }
% \maketitle              

%%
%% The "author" command and its associated commands are used to define the authors and their affiliations.
% \author{Youhuan Li}
% \affiliation{%
%  \institution{Hunan University}
% %  \streetaddress{P.O. Box 1212}
% %  \postcode{43017-6221}
% }
% \email{liyouhuan@hnu.edu.cn}

% \author{Lei Zou} %$^\dagger$
% \affiliation{%
%  \institution{Peking University}
% }
% \email{zoulei@pku.edu.cn}
% \authornote{Lei Zou is the corresponding author.}

% \linepenalty=1000
% \captionsetup[table]{skip=5pt}
\maketitle 
% \vspace{-15pt}
\begin{abstract}
% 单查询问题
Finding $k$ disjoint paths ($k$DP) is a fundamental problem in graph analysis. 
    For vertices $s$ and $t$, paths from $s$ to $t$ are said to be \textit{disjoint} 
        if any two of them share no common vertex except $s$ and $t$. 
In practice, disjoint paths are widely applied in network routing and transportation.
% 多查询问题
In these scenarios, \textit{multiple} $k$DP queries are often issued simultaneously, 
    necessitating efficient batch processing.
This motivates the study of \textit{batch} $k$DP query processing (\texttt{batch-$k$DP}).
% 批量路径枚举方法+disjoint 
A straightforward approach to \texttt{batch-$k$DP} extends batch simple-path enumeration with disjointness checks.
    But this suffers from factorial computational complexity. 
% 可以不用这么高复杂度
An alternative approach leverages single-query algorithms that avoid this by 
    replacing the graph with a converted version. 
% 共享空间
However, handling each query independently misses opportunities for shared computation.
% can lead to redundant computations and storage needs.
% 我们的方法
To overcome these limitations, we propose \texttt{ShareDP}, an algorithm for \texttt{batch-$k$DP} 
    that \texttt{share}s the computation and storage across $k$\texttt{DP}s. 
    % that \texttt{share}s these redundancies across $k$\texttt{DP}s. 
\texttt{ShareDP} merges converted graphs into a shared structure, then shares the traversals and operations from different queries within this structure. 
Extensive experiments on 12 real-world datasets confirm the superiority of \texttt{ShareDP} over comparative approaches.

\end{abstract}
% \vspace{-10pt}

%!TEX root = main.tex
\section{Introduction}\label{sec:introduction}
\begin{figure}[h!]
\newcommand{\mylinewidth}{\linewidth}
\centering
    \begin{subfigure}[t]{0.35\mylinewidth}
        \centering
        % \resizebox{\mylinewidth}{!}
        {\includegraphics[width=\linewidth]{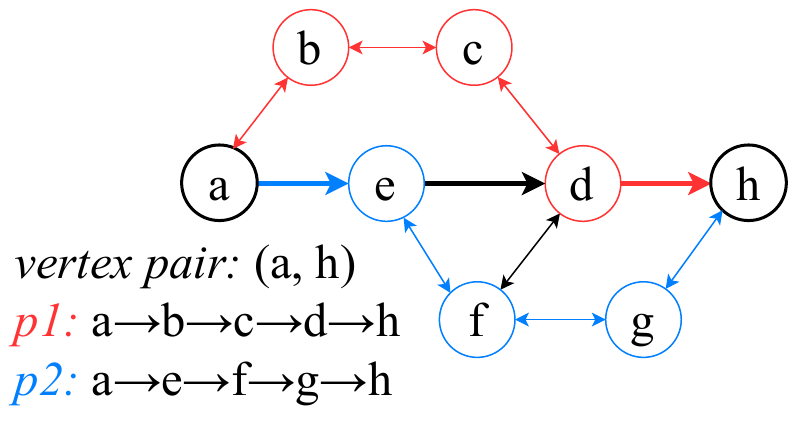}}
        \caption{Disjoint paths for $(a, h)$.}
        \label{fig:g}
    \end{subfigure}
    \begin{subfigure}[t]{0.6\mylinewidth}
        \centering
        % \resizebox{\mylinewidth}{!}
        {\includegraphics[width=\linewidth]{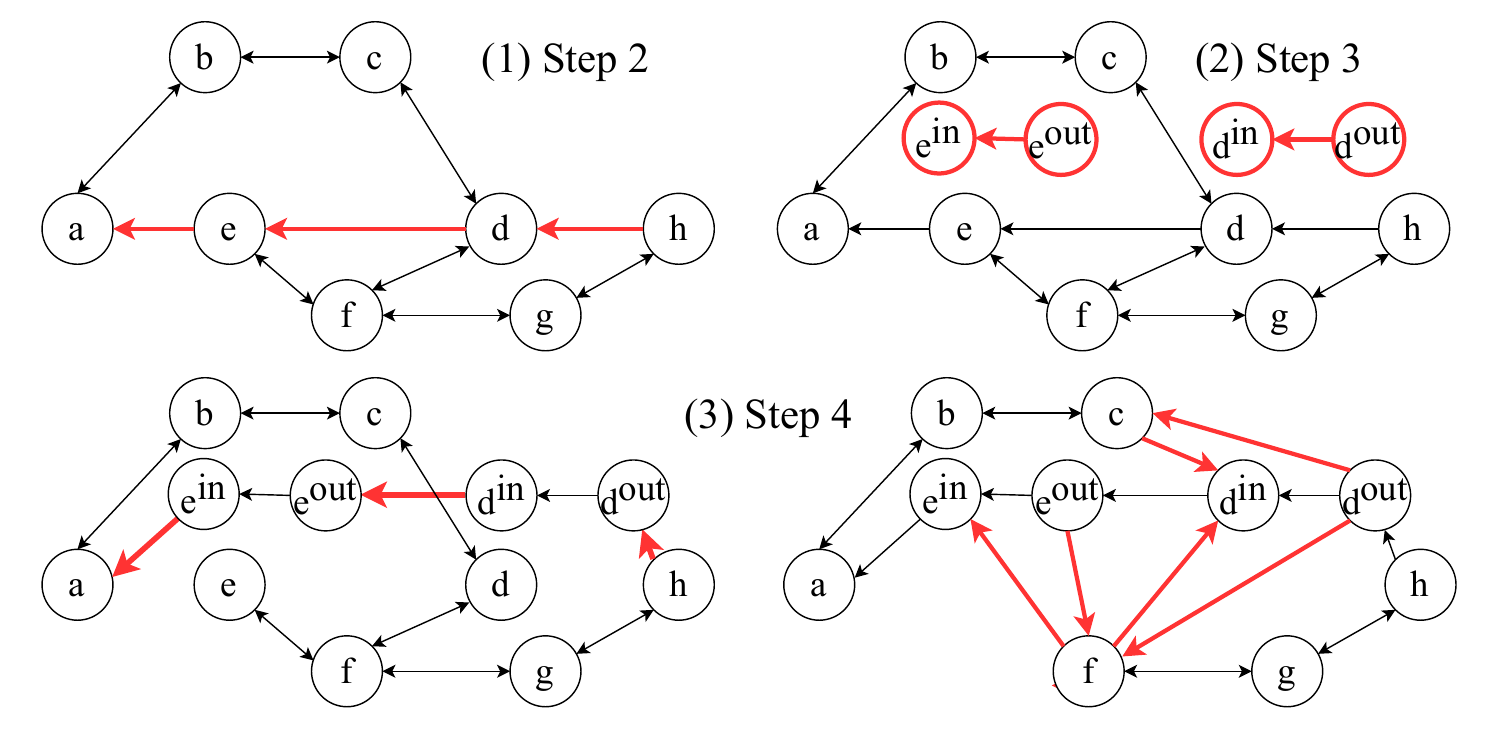}}
        \caption{Split-graph with $P= \{p_1=\{$a$, $e$, $d$, $h$\}\}$.}
        \label{fig:eg_split}
    \end{subfigure}
    \caption{Examples of disjoint paths and split-graph.}
    % \label{fig:fg_share_intuition}
\end{figure}

% 单查询问题和应用
Finding $k$ disjoint paths ($k$DP) is a fundamental challenge in graph analysis~\cite{baseline_moreverbose,baselineOnlySplitP1,baseline1step2, Penalty,dissimilarity_topk1,SCB}. 
Given a source and a target vertex, $k$DP identifies $k$ paths that do not share any vertices except for the source and target\footnote{We focus on vertex-disjoint paths, as edge-disjoint path finding problem can be reduced to vertex-disjoint version in polynomial time~\cite{edgedisjoint2vertexdisjoint}.}. 
Fig.~\ref{fig:g} illustrates an example with disjoint paths $p_1$ (red) and $p_2$ (blue).
Applications of $k$DP include cyber-security, network fault tolerance, and load balancing~\cite{2009BRTree,cyber_secure, Penalty,dissimilarity_topk1,SCB}.

% 多查询问题
In practice, multiple $k$DP queries are often processed in batches. 
For example, as communication networks scale with the Internet, numerous routing queries are generated quickly, necessitating high-throughput processing.
This paper focuses on the problem of batch $k$DP query processing (\texttt{batch-$k$DP}), where given a graph $G$ and a set of $k$DP queries $Q$, returns $k$ disjoint paths for each query in $Q$.

% 相关工作：我们工作的定位
Existing studies on $k$DP can be categorized into \emph{single-query}~\cite{baseline_moreverbose,baselineOnlySplitP1,baseline1step2,Penalty,dissimilarity_topk1,SCB} and \emph{single-source}~\cite{2009BRTree,four_independent_spanning,IST_survey}. 
(1) \emph{Single-query} methods include \emph{flow-augmenting path-based}~\cite{baseline_moreverbose,baselineOnlySplitP1,baseline1step2} and \emph{dissimilar path-based}~\cite{Penalty,dissimilarity_topk1,SCB}. 
The latter extends simple path enumeration by incorporating disjointness constraints, but suffers from factorial time complexity (see Sec.~\ref{sec:related_kdp}).
In contrast, the former achieves linear time complexity using \emph{split-graphs}~\cite{baseline_moreverbose,baseline1step2,baselineOnlySplitP1}, where finding disjoint paths reduces to identifying flow-augmenting paths in the split-graphs. 
(2) \emph{Single-source} $k$DP seeks disjoint paths from a source to all other vertices, with current studies primarily theoretical and limited to $k \leq 4$~\cite{IST_survey}. 
(3) To our knowledge, this paper is the first to address \texttt{batch-$k$DP} with general $k$.
% for large graphs with general $k$.

% 解决batch-kDP问题的baseline & 多查询共享机会
For \texttt{batch-$k$DP}, we can adapt methods from single-query $k$DP: 
(1) \emph{Dissimilar path-based} methods can extend batch path enumeration algorithms~\cite{BatchEnum} but also face factorial complexity. 
(2) \emph{Flow-augmenting path-based} methods handle each query in linear time but require independent split-graphs for each query, missing opportunities for shared computation 
% (also observed in empirical studies, see Sec.~\ref{sec:approach_algo}). 
% Our empirical studies reveal significant potential for shared computation (see Sec.~\ref{sec:approach_algo}).

% 用不了已有的多查询技术
This leads to the question: can we apply existing batch-processing techniques from other problems? 
Unfortunately, current techniques are designed for queries in the same graph~\cite{BatchEnum, tods21subgraphiso, icde20shortestpath, msbfs}, while our problem involves different $k$DPs executed over different graphs (split-graphs). 
The first step is to unify the split-graphs, which is not covered in the literature. 
Thus, we propose \texttt{ShareDP}, a batch-$k$DP algorithm that aims to \texttt{share} common computations and storage across a batch of $k$DP queries. 
The overall framework of \texttt{ShareDP} is illustrated in Fig.~\ref{fig:shareDP_overview} (explained in Sec.~\ref{sec:approach_overview}).

% 我们的设计
\texttt{ShareDP} incorporates the following key strategies:
(1) We consolidate individual split-graphs into a unified structure, represented implicitly via result sets, simplifying construction and updates.
(2) Using the merged split-graph, we enable concurrent $k$DP query processing by consolidating traversals and operations leveraging tagged data structures, reducing redundancy and improving efficiency.
% (3) Related queries are grouped into batches for more efficient processing.
In summary, our contributions are:

\begin{itemize}[]
% \vspace{-5pt}
\item \emph{Merged Split-Graph Representation.} 
The \texttt{ShareDP} framework introduces a novel merged split-graph representation that consolidates individual split-graphs into a unified structure, enabling dynamic sharing of traversals.

\item \emph{Optimized Path-Finding with Shared Traversals.} 
The framework combines traversals across multiple queries, consolidating common operations into single steps, reducing redundancy and improving efficiency.

\item \emph{Proven Efficiency and Scalability.}  
Extensive evaluations on 12 real-world datasets demonstrate that \texttt{ShareDP} consistently achieves the lowest runtime across various \( k \) settings, confirming its efficiency and scalability.
\end{itemize}

% 我们的方法
% 我们主要设计了什么
% To achieve shared computation, we address three challenges:
% (1) Merging split-graphs into a shared structure for unified traversals. 
% (2) Efficiently representing, constructing, and updating the merged split-graph. 
% (3) Executing $k$DP queries based on the merged split-graph to share common computations.

% To achieve shared computation, we need to solve the following three challenges:
%     (1) How to merge the split-graphs to a shared structure 
%         so that traversals from different queries can be executed in one unified structure? 
%     (2) How to represent the merged split-graph, and how to efficiently construct and update it? 
%     (3) Based on the merged split-graph, how to execute the $k$DP queries such that common computations can be shared?

% 我们的设计
% To address these challenges, we propose the \texttt{ShareDP} framework with the following key strategies:
\section{Preliminaries}\label{sec:problemdef}

% A directed graph is a $2$-tuple $G=(V, E)$, where $V$ is the set of vertices and $E$ is the set of edges. 
We use $p$ for a path and $P$ for a set of paths.
A vertex $v$ is an \emph{intermediate} vertex in path $p$ if it is neither the starting nor the ending vertex. If $v$ is intermediate in any path in $P$, it is called \emph{$P$-inner}.
A path $p$ is simple if no vertex appears more than once.
$V(P)$ and $E(P)$ refer to the set of vertices and edges that appear in at least one path in $P$.
While the focus here is on directed graphs, the method can be easily adapted to undirected graphs.

\textbf{Disjoint Paths}
Given a graph $G = (V, E)$ and two simple paths $p_1$ and $p_2$ from $s$ to $t$, $p_1$ and $p_2$ are \emph{disjoint} if they share no common vertex except $s$ or $t$.

\textbf{Problem Statement}
Given a graph $G = (V, E)$, a parameter $k$, and a set of $k$DP queries $Q = \{q_0, q_1, ..., q_{\omega}\}$ where each query $q_i$ is a vertex pair $(s_i, t_i)$ ($i = 1, ..., \omega$), 
\texttt{batch-$k$DP} is to find $k$ disjoint paths for each query $(s_i, t_i) \in Q$.

% \textbf{Example:}
% Consider the graph in Fig.~\ref{fig:g} with \( k = 2 \) and a set of queries \( Q = \{q_0:(a, h), q_1:(b, f), q_2:(c, f)\} \). 
% For $q_0$, two disjoint paths, $p_1$ (red) and $p_2$ (blue), can be identified. Solutions for the remaining queries are found similarly.

\section{Related Work}
\label{sec:relatedwork}

% \vspace{-5pt}
\subsection{$k$DP problems}\label{sec:related_kdp}
\textbf{Single-query.} 
Approaches to single-query $k$DP can be divided into \emph{flow-augmenting path-based methods}~\cite{baseline_moreverbose, baselineOnlySplitP1, baseline1step2} and \emph{dissimilar path-based methods}~\cite{Penalty, dissimilarity_topk1, SCB}.
Flow-augmenting methods use \emph{split-graphs}~\cite{baseline_moreverbose, baseline1step2}, achieving linear time complexity (Sec.~\ref{sec:splitgraph}).
Dissimilar path-based methods, which define dissimilarity as disjointness, include \emph{penalty-based}, \emph{dissimilarity-based}, and \emph{plateaus-based} methods~\cite{Penalty, SCB}.
Plateaus-based methods, however, may fail to find solutions, as they depend on shared branches between two shortest spanning trees, which may not present in all $k$DP solutions. Thus, we exclude them from comparison.

Penalty-based and dissimilarity-based methods reduce to path enumeration with disjointness constraints, either by \emph{(1)} marking vertices in previous paths as inaccessible or by \emph{(2)}  verifying disjointness before adding new paths. However, both lead to factorial time complexity in the worst case.
For instance, if a path $p$ is not part of any solution, all computations involving $p$ in the result set will fail to yield a solution. 
    Consequently, alternative path orderings must be attempted (e.g., adding $p_1$ first, then $p_2$, and so on).
    In the worst case, every possible path ordering must be evaluated, leading to factorial time complexity.
We may also \emph{(3)} first identify all paths and then select a subset of $k$ disjoint paths, 
    but this also faces the path-ordering challenge, resulting in the same factorial complexity.
Our experiments (Sec.~\ref{sec:experiment_time}) show these methods time out on large graphs.

\emph{$k$ shortest dissimilar path finding}~\cite{dissimilarity_topk1} can solve $k$DP but is slower due to path enumeration in ascending length order, as noted in~\cite{SCB}. Thus, we exclude them from comparison.

\textbf{Single-source.} 
The problem of finding $k$ independent spanning trees~\cite{2009BRTree, four_independent_spanning, IST_survey} is related to $k$DP, where paths in different spanning trees are disjoint. While this provides solutions for vertex pairs formed by the root and other vertices, the problem remains an open challenge for $k>4$. For $k\le 4$, this method was over 10 times slower than our approach in our experiments (Sec.~\ref{sec:experiment_time}).

To our knowledge, no prior work addresses the \texttt{batch-$k$DP} problem.

\subsection{Batch Query Processing} \label{sec:related_batch}
Batch query processing has been studied for other graph problems~\cite{BatchEnum, tods21subgraphiso, icde20shortestpath, msbfs}, focusing on queries within the same graph.
\cite{BatchEnum} explores batch simple path enumeration, leveraging shared computations and caching results. \cite{tods21subgraphiso} investigates batch subgraph isomorphism with a unified join plan. \cite{icde20shortestpath} implements caching and query decomposition for shortest path queries. Batch BFS queries~\cite{msbfs} exploit overlaps among frontiers across concurrent BFS runs.
In contrast, our problem involves different $k$DP queries on different graphs (split-graphs), requiring unification of these graphs, which remains unexplored in the literature.

\section{Baseline} \label{sec:splitgraph}

The baseline method for batch-$k$DP solves each query using flow-augmenting path-based methods, which rely on the concept of \textit{split-graphs}~\cite{baseline_moreverbose, baseline1step2, baselineOnlySplitP1}. 
% For each query, paths are iteratively found in a split-graph, which is updated after each iteration.
% A split-graph is constructed by two transformations of the original graph:
% (1) reversing result-set paths, simulating flow-augmentation, and 
% (2) splitting vertices within these paths, giving rise to the name ``split-graph."

\textbf{Definition: Split-Graph~\cite{baselineOnlySplitP1}} 
Given a graph \( G = (V, E) \) and a set \( P \) of disjoint paths from \( s \) to \( t \), the split-graph \( \iG_{G,P} = (\iV_{G,P}, \iE_{G,P}) \) is constructed as follows:
(1) Initializing \( \iV_{G,P} = V \) and \( \iE_{G,P} = E \).
(2) For each edge in \( E(P) \), reversing the corresponding edge in \( \iE_{G,P} \).
(3) Splitting vertices \(v \in V(P) \setminus \{s, t\}\) into \(v^{in}\) and \(v^{out}\), and connecting them accordingly.
(4) Replacing edges in \(\iE_{G,P}\) with updated vertex connections, preserving incoming and outgoing edges.

% \textbf{Example}: 
% Fig.~\ref{fig:eg_split} shows the split-graph construction for the graph \( G \) in Fig.~\ref{fig:g} with $P= \{p_1=\{a, e, d, h\}\}$. Changes are shown in red.

% \vspace{-10pt}
% \vspace{-5pt}

% 删除 begin
Given a graph \( G \) and vertices \( s \) and \( t \), the algorithm proceeds as follows:
% (1) Initialize \( P = \emptyset \) and \( \iG_{G,P} = G \).
% (2) Find the first path \( p_1 \) using a path-finding algorithm (e.g., BFS) in \( \iG_{G,P} \) and update \( \iG_{G,P} \).
% (3) Find the second path \( p_2 \), update found paths following an approach similar to augmenting paths in the maximum flow problem~\cite{baseline_moreverbose}, then update \( \iG_{G,P} \). More paths are found in a similar manner.
(1) Initialize $P = \emptyset$ and $\iG_{G, P} = G$.
(2) Find the first path $p_1$ in $\iG_{G, P}$ using any path-finding algorithm (e.g., BFS), forming $P_1 = \{p_1\}$, and update $\iG_{G, P}$ to $\iG_{G, P_1}$.
(3) Search for $p_2$ in $\iG_{G, P_1}$, yielding $P_2 = \{p_1, p_2\}$, and adjust $P_2$ following an approach similar to augmenting flows~\cite{baseline_moreverbose}.
Then update $\iG_{G, P_1}$ to $\iG_{G, P_2}$.
(4) Search for $p_3$ in $\iG_{G, P_2}$. More paths are found in a similar manner.
% 删除 end
% 先压缩实验，然后方法
% 方法能不能，有个overview就差不多呢
%!TEX root = main.tex
\section{ShareDP} \label{sec:approach_algo}

Given a set of \texttt{$k$DP} queries, the key improvement over the baseline in our approach lies in sharing computations across queries. 
We demonstrate this with the indochina-2004 dataset (statistics in Tab.~\ref{tab:datasets}). 
In both the first and last iterations (i.e., finding the first and $k$-th disjoint paths), exploration of over 60\% of the vertices are shared across levels, and in more than half of the levels, over 80\% are shared. 
This illustrates the potential for shared exploration in the batch-$k$DP problem. Based on this, we propose the \texttt{ShareDP} algorithm.

% \vspace{-10pt}
\subsection{Algorithm Overview} \label{sec:approach_overview}

\begin{figure}
    \centering
    % \resizebox{\linewidth}{!}
    {\includegraphics[width=\linewidth]{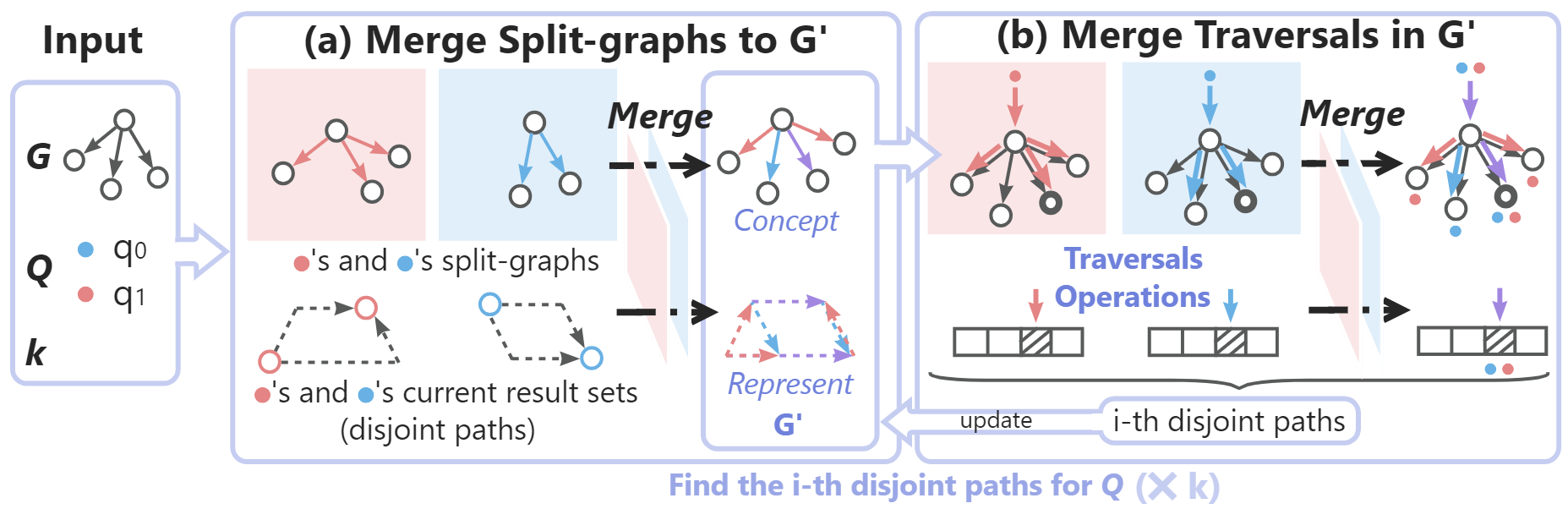}}
    % \vspace{-10pt}
        \caption{Framework of \texttt{ShareDP}. Colored balls (e.g., the blue ball for query \( q_0 \)) represent queries. 
	Each iteration identifies a path for each query 
    by conducting a combined search on a merged split-graph $G'$.
        % enabling the sharing of traversals and operations.
See Sec.~\ref{sec:approach_overview} for details.}
    \label{fig:shareDP_overview}
\end{figure}

The \texttt{ShareDP} framework (Fig.~\ref{fig:shareDP_overview}) takes as input a graph \( G \), a set of $k$DP queries \( Q \), and an integer \( k \). 
In each iteration, the algorithm identifies one path for each query. 
In the \( i \)-th iteration, the algorithm operates on a merged split-graph \( G' \) , 
	conceptually the union of individual split-graphs for each query (top half of Fig.~\ref{fig:shareDP_overview}(a)). 
	\( G' \) is represented implicitly using the current result sets of all queries (bottom half of Fig.~\ref{fig:shareDP_overview}(a)). 
During path-finding, traversals are combined across queries, 
	conceptually visualized as collapsing multiple planes where the traversal for each query flows into a unified plane (top half of Fig.~\ref{fig:shareDP_overview}(b)).
	Common operations are consolidated into a merged step by tagging data structures with query sets (bottom half of Fig.~\ref{fig:shareDP_overview}(b)).  
Paths identified in each iteration update the result sets, which in turn update \( G' \) for the next iteration (bottom of Fig.~\ref{fig:shareDP_overview}(b)). 

% \vspace{-15pt}
% \vspace{-10pt}

% \vspace{-10pt}
\subsection{Algorithm Details} \label{sec:approach_details}
% \subsection{Merge the Split-Graphs} \label{sec:approach_topo}

\emph{Merge the Split-Graphs.} We merge the split-graphs from the perspective of a vertex \( v \), focusing on its out-neighbors (in-neighbors analysis is symmetric). 
They are derived from the original graph’s out-neighbors, reversed edges, and vertex splitting (Sec.~\ref{sec:splitgraph}). 
To manage edge reversals and vertex positions, we define \( nexthops \), \( prehops \), \( isPinner \), \( isS \), and \( isT \), which also represent the current result sets $\{P\}$. 
Specifically, 
    \( prehops_{u,v} \) represents queries where \( v \) is \( u \)'s $prehop$; 
    \( isPinner_v \), \( isS_v \), and \( isT_v \) represent the set of $k$DP queries where \( v \) is \( P \)-inner, \( s \), and \( t \), respectively.
With these definitions, given a query set \( B \) and a vertex \( v \), the procedure for acquiring \( v \)'s out-neighbors is shown in Alg.~\ref{alg:get_nbr}. 
The merging process uses the result sets \(\{P\}\) from each query \( q \in B \) to determine neighbors in the merged split-graph \( G' \), simplifying neighbor retrieval and improving efficiency compared to constructing a supergraph (Sec.~\ref{sec:experiment_opt}).

\begin{algorithm}[t]
% \scriptsize
    \caption{GetOutNeighbors}
    \label{alg:get_nbr}
    \KwIn{
        Vertex $v$, set of $k$DPs $B$, original graph $G$, \{$P$\}
        % \textit{\color{gray} // \{$P$\}: represented by $prehops$, $nexthops$, $isPinner$, $isS$ and $isT$.}
    }
    \KwOut{\{($u$, $B'$)\} \textit{\color{gray} // $u$: a neighbor of $v$, $B'$: corresponding $k$DP subset.}}
    $S = \emptyset$ \\ % newline
    \uIf{$v$ is $new_{in}$ \textit{\color{gray} // $v$ is $v_{in}$.}}
    {
        \ForEach{$u$ in $prehops_v$ \textit{\color{gray} // Reversed edges: $v$'s prehops.}}
        {
            Add ($u$, $B \cap prehops_{v, u}$) to $S$ 
            % \textit{\color{gray} // $k$DPs where $u$ is $v$'s $prehop$.}
        }
    }
    \Else(\textit{\color{gray} // $v$ is $v_{out}$ or $v$.})
    {
        \ForEach{$u$ in out-neighbors of $G$}
       {
            $haveset = B \setminus nexthops_{v, u}$ \textit{\color{gray} // Excluding reversed edges.}\\
            % \textit{\color{gray} // Edges not in any $P$ of $haveset$.} \\
            Add ($u^{in}$, $haveset \cap isPinner_u$) to $S$ \textit{\color{gray} // Edges to $P$-$inner$ vertices are redirected to $new^{in}$.} \\
            Add ($u$, $haveset \setminus isPinner_u$) to $S$ 
            \textit{\color{gray} // Otherwise no redirecting.} 
            % \textit{\color{gray} // Edges to non-$P$-$inner$ vertices.} 
            \\
        }
        Add ($v^{in}$, $B \cap isPinner_v$) to $S$ 
        \textit{\color{gray} // The edge from $v^{out}$ to $v^{in}$}
        % \textit{\color{gray} // $v^{out}$ has an out-going edge to $v^{in}$}
        \\
    }
    \Return $S$\\
    % \vspace{-2pt}
\end{algorithm}

% \subsection{Combine Traversals in the Merged Split-Graph} \label{sec:approach_algorithm_list}
\emph{Combine Traversals.}
% To get a picture of how the traversals get merged in the merged split-graph $G'$, assume that each $k$DP runs separately.
% Traversals start from their source vertices, and may meet in certain regions. Then they proceed together until they diverge.
The ShareDP algorithm combine the traversals by tagging data structures with query sets. 
While any path-finding algorithm could be used, we employ bidirectional BFS as an example here.
% The algorithm iteratively finds disjoint paths. 
% For each $i$-th iteration, each query is initialized with its source vertex and a $k$DP set. 
% The algorithm expands vertices in the frontier until all $k$DPs find their $i$-th path or the queue empties. 
When a frontier vertex $v$ expands in the forward search (Alg.~\ref{alg:expandFrontier}, 
backward search proceeds similarly), its $k$DP set is divided into subsets, and each neighbor is processed only once for the $k$DPs in that subset.
If frontiers share the same neighbor, their $k$DP sets are merged, enabling continued shared traversal.
The complete algorithm of \texttt{ShareDP} is provided in the appendix (arXiv).

\begin{algorithm}[h]
% \scriptsize
    \caption{ForwardExpandFrontier}
    \label{alg:expandFrontier}
    \KwIn{Frontier (vertex $v$, queries $B$), original graph $G$, \{$P$\}}
    % \textbf{InOut:} $undone$, $s$-$seen$, $s$-$queue$, $s$-$nextqueue$, $joint$, $pred$, $t$-$seen$ \textit{\color{gray} // Data structures for search and path record.} \\
    \textbf{InOut:} Data structures for searching and path recording. \\
    % \textit{\color{gray}/* Forward search. Backward search proceeds similarly. */} \\
    {
        $B = B \setminus undone$ \textit{\color{gray}// Skip queries that have already found the $i$th path.} \\
        \ForEach{($u$, $B'$) $\in$ GetOutNeighbors($v$, $B$, $G$, \{$P$\}) \textit{\color{gray}// See Alg.~\ref{alg:get_nbr}.}} 
        {
            $D = B' \setminus s$-$seen_u$ \textit{\color{gray}// Exclude queries that have already visited $u$.} \\
            $s$-$seen_u \cup= D$; $pred_{u, v} \cup= D$ \textit{\color{gray}// Mark visited and record predecessor.} \\
            $meet = D \cap t$-$seen_u$ 
            \textit{\color{gray}// The forward and backward searches meet.} \\
            % \textit{\color{gray}// Found a path} \\
            $joint_u \cup= meet$; $undone \setminus= meet$ \textit{\color{gray}// Mark queries as completed.} \\
            $s$-$nextqueue_u \cup= (D \setminus meet)$ 
            \textit{\color{gray}// Add to the queue.} \\
        }
    }
    
    % \vspace{-2pt}
\end{algorithm}

% % \vspace{-5pt}

%!TEX root = main.tex
% \vspace{-3pt}
\section{Experimental Evaluation}\label{sec:experiment}

We implemented all algorithms in C++ and tested on a Ubuntu machine with 512 GB RAM and Intel(R) Xeon(R) Platinum 8352V 2.10GHz CPU. 

% \vspace{-5pt}
\subsection{Experimental Setup}\label{sec:experiment_setup}
% We implemented all algorithms in C++ and tested on a Ubuntu machine with 512 GB RAM and Intel(R) Xeon(R) Platinum 8352V 2.10GHz CPU. 

% \subsubsection{Algorithms}\label{sec:experiment_algo}
{\textbf{Algorithms.}}
% We compare the following algorithms:
1) \emph{{\texttt{ShareDP}}}: Our approach using bidirectional BFS (Sec.~\ref{sec:approach_details}).
2) \emph{{\texttt{ShareDP-}}}: \texttt{ShareDP} with supergraph representation.
3) \emph{{\texttt{maxflow}}}~\cite{baselineOnlySplitP1}: Baseline method (Sec.~\ref{sec:splitgraph}).
4) \emph{{\texttt{BatchEnum}}}~\cite{BatchEnum}: State-of-the-art path enumeration with adaptations (3) for our problem (Sec.~\ref{sec:related_kdp}).\footnote{BatchEnum reuses path enumeration results across queries, requiring complete enumeration, which prevents adaptations (1) and (2). }
5) \emph{{\texttt{SCB+}}}~\cite{SCB}: State-of-the-art dissimilar path finding method (adaptation (2)).
6) \emph{{\texttt{Penalty}}}~\cite{Penalty}: Another dissimilar path finding method (adaptation (1)).
7) \emph{\texttt{IST}}~\cite{2009BRTree}: Single-source method for \( k = 2 \) (case for $k > 4$ is open challenge, see Sec.~\ref{sec:related_kdp}).

% For \texttt{ShareDP} and \texttt{ShareDP-}, we split the query set into equal-sized batches and run them sequentially, balancing shared computation and memory overhead. 
% Specifically, batch sizes of $128$ ($64$) were used for large (small) graphs.

% \subsubsection{Datasets And Queries}\label{sec:experiment_dataset}
{\textbf{Datasets and Queries.}}
We evaluate on 12 real-world datasets (Tab.~\ref{tab:datasets}), sourced from \href{https://snap.stanford.edu/}{SNAP}, \href{http://networkrepository.com/networks.php}{NetworkRepository}, and \href{https://law.di.unimi.it/index.php}{LAW}. 
These include the datasets used in \cite{SCB} (first 6 rows), supplemented by 6 larger datasets for improved comparison (last 6 rows).
% queries
For each graph, we generate 1000 vertex pairs with \( k \)DP solutions, starting from \( k \)=\( 50 \) down to \( k \)=\( 2 \). 
Candidate pairs are selected based on vertex degree \( \geq k \). 
If fewer than 20\% succeed, \( k \) is reduced. The maximum \( k \) is termed \( k_{max} \). Algorithms are evaluated on 4–5 \( k \)s per graph, based on \( k_{max} \).

% % % \vspace{-10pt}
% \begin{table}[!h]
%     \centering
%     \caption{Properties and evaluated $k$ values of datasets}
%     % \resizebox{\linewidth}{!}
%     {
%         \input{tab/tab_datasets}
%     }
%     \label{tab:datasets}
% \end{table}
% % % \vspace{-10pt}

% \vspace{-10pt}
\subsection{Experimental Result}\label{sec:experiment_result}
\subsubsection{Comparing Algorithms when varying $k$}\label{sec:experiment_time}

\begin{figure}
    \centering
    \includegraphics[width=\linewidth]{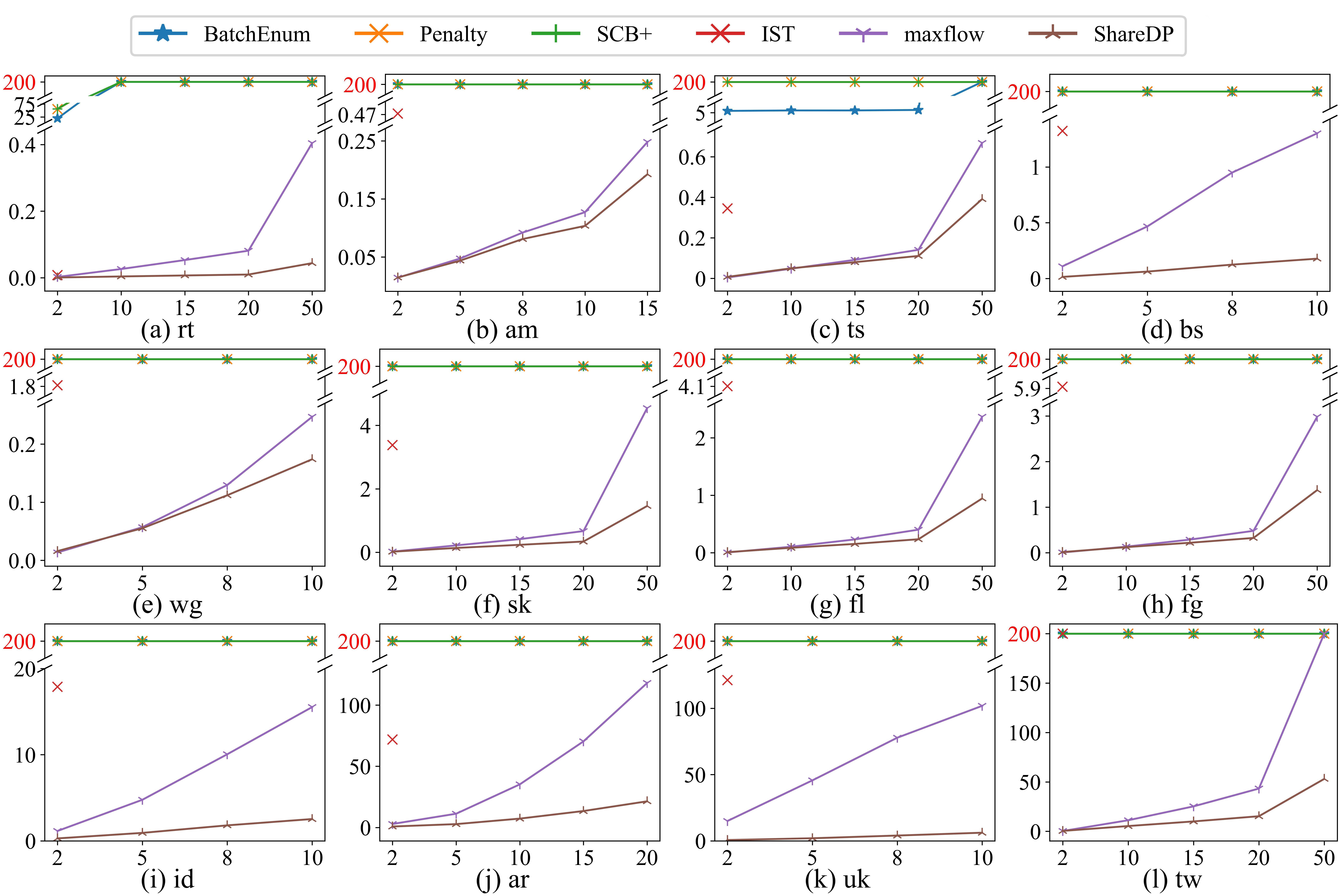}
    % \vspace{-10pt}
    \caption{Running time when varying $k$.}
    \label{fig:running_time}
\end{figure}

We evaluated different algorithms across various $k$ values (Fig.~\ref{fig:running_time}). 
The y-axis shows average runtime per query (seconds), and the x-axis represents $k$. 
Queries exceeding 200 seconds were terminated and recorded as such. 
% For \texttt{ShareDP++}, batch sizes of $128$ and $64$ were used for large and small graphs, respectively, to balance shared computation with overhead.

\texttt{ShareDP} (brown) consistently outperformed others with the lowest runtime. 
Most dissimilarity-based methods failed to complete for nearly all $k$s, 
% except for small graphs (rt, ts), where they also failed for larger $k$. 
When they did run, they were more than 10 times slower than flow-based methods (\texttt{maxflow} and \texttt{ShareDP}). 
\texttt{IST} was at least 10 times slower for the only \( k \) it handled (\( k \)=2).

% \texttt{maxflow} (purple line) showed faster performance than the slower methods but grew quickly with $k$. 
\texttt{ShareDP}'s advantage over \texttt{maxflow} increased as $k$ grew. For graphs with fewer disjoint paths (e.g., id and uk), \texttt{ShareDP} significantly outperformed \texttt{maxflow}. For graphs with more disjoint paths (e.g., sk and tw), the advantage of \texttt{ShareDP} was more pronounced at larger $k$, as finding a small number of disjoint paths in these graphs is easy, reducing the relative advantage of \texttt{ShareDP} in simpler cases. For similar reasons, \texttt{ShareDP} showed a greater advantage on larger graphs (e.g., the last row). These results highlight the efficiency and scalability of \texttt{ShareDP}.

% % \vspace{-20pt}
% \begin{figure}
%     \centering
%     \includegraphics[width=\linewidth]{pic//exp/k_running_time}
%     % \vspace{-10pt}
%     \caption{Running time when varying $k$.}
%     \label{fig:running_time}
% \end{figure}
% % % \vspace{-25pt}
% % \vspace{-20pt}

% % \vspace{-10pt}
\begin{table}[!h]
    \centering
    \caption{Properties and evaluated $k$ values of datasets}
    % \resizebox{\linewidth}{!}
    {
        % \begin{small}
% \scriptsize
% \vspace{-5pt}
\begin{tabular}{|l|l|l|l|l|l|l|l|}
\hline
    {\bfseries Name} & \multicolumn{1} {c|}
    {\bfseries Dataset} & \multicolumn{1} {c|}
    {\bfseries $\left|V\right|$} & \multicolumn{1} {c|} 
    {\bfseries $\left|E\right|$} & \multicolumn{1} {c|} 
    {\bfseries Type} & \multicolumn{1} {c|} 
    {\bfseries D} & \multicolumn{1} {c|} 
    {\bfseries $k_{max}$} & \multicolumn{1} {c|} 
    {\bfseries $k$} \\
\hline
rt & reactome & 6.3K & 147K & Biology & 24 & 50 & 2,10,15,20,50 \\
am & amazon & 334K & 925K & Web & 44 & 15 & 2,5,8,10,15 \\
ts & twitter-social & 465K & 834K & social  & 8 & 50 & 2,10,15,20,50 \\
bs & berkstan & 685K & 7M & Web & 208 & 10 & 2,5,8,10 \\
wg & web-google & 875K & 5M & Web & 24 & 10 & 2,5,8,10 \\
sk & skitter & 1.6M & 11M & infrastructure & 31 & 50 & 2,10,15,20,50 \\
fl & flickr-links & 1.7M & 16M & Social & 14 & 50 & 2,10,15,20,50 \\
fg & flickr-growth & 2.3M & 33.1M & Social & 13 & 50 & 2,10,15,20,50 \\
id & indochina-2004 & 7.4M & 194M & Web & 25 & 10 & 2,5,8,10 \\
ar & arabic-2005 & 22.7M & 640M & Web & 29 & 20 & 2,5,10,15,20 \\
uk & uk-2005 & 39.5M & 1.9B & Web & 22 & 10 & 2,5,8,10 \\
tw & twitter-2010 & 42M & 1.5B & Social & 16 & 50 & 2,10,15,20,50 \\
\hline
\end{tabular}
% \end{small}

    }
    \label{tab:datasets}
\end{table}
% % \vspace{-10pt}

% % 删除 begin
\subsubsection{Effect of the number of $k$DPs}\label{sec:experiment_instance_number}

\begin{figure}
    \centering
    \includegraphics[width=\linewidth]{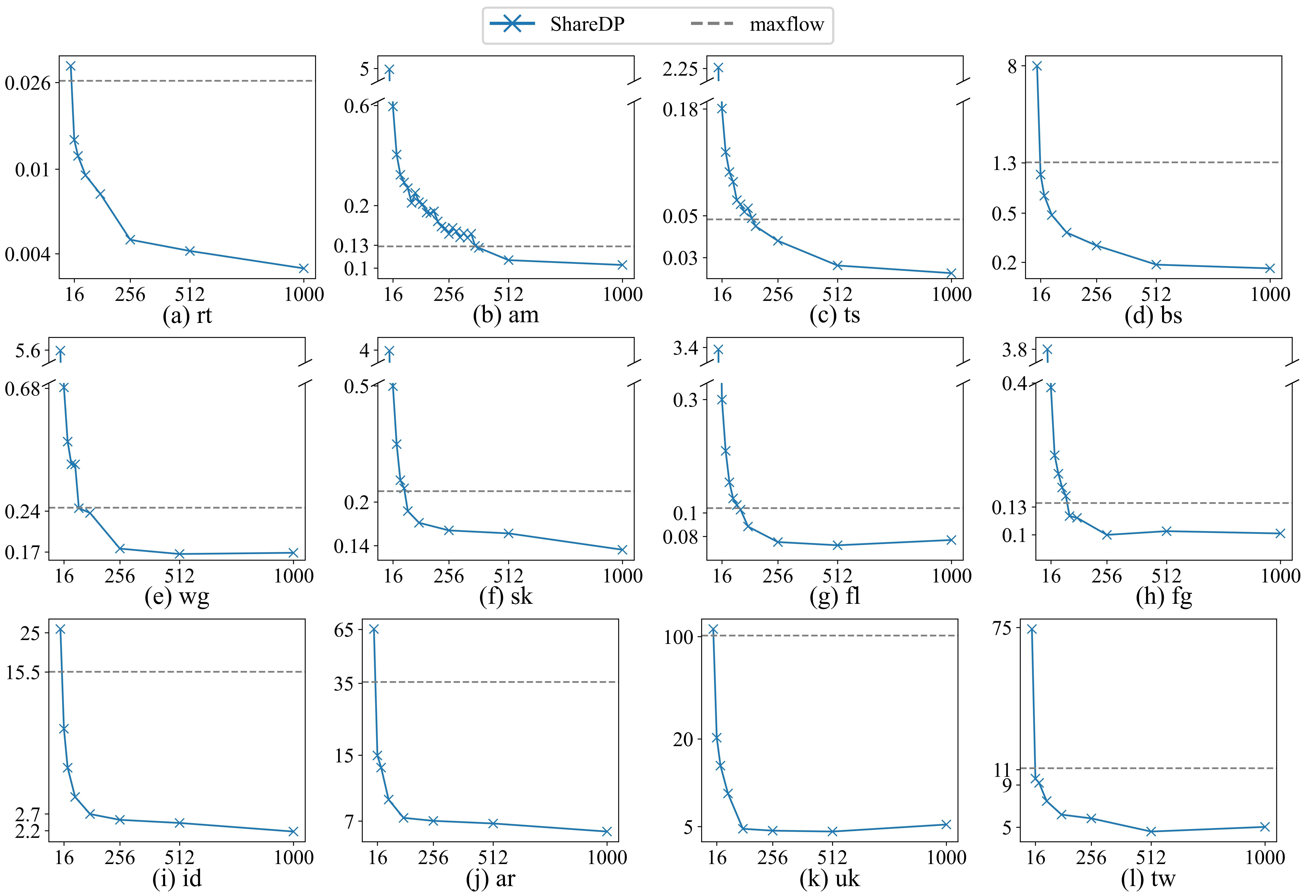}
    % \vspace{-10pt}
    \caption{Running time when varying the number of $k$DPs with $k$=$10$.}
    \label{fig:vpcnt_impact}
\end{figure}

% The goal of this paper is to efficiently execute a large number of $k$DPs.  
We analyzed performance as the number of $k$DPs ($|Q|$) varied from 1 to 1000 with $k$=$10$ across all datasets (Fig.~\ref{fig:vpcnt_impact}).  
% For each $|Q|$, we generated $max\{256/|Q|, 1\}$ samples by random selection from the query set in Sec.~\ref{sec:experiment_setup}, averaging the runtime per sample. 
The y-axis shows average runtime per query (seconds), and the x-axis shows $|Q|$.
As shown in Fig.~\ref{fig:vpcnt_impact}, \texttt{ShareDP}'s runtime decreases as $|Q|$ increases, outperforming \texttt{maxflow} at small $|Q|$.  
This is due to shared computation when more $k$DPs are processed together.  
The improvement slows with larger $|Q|$ as the average shared computation stabilizes, reaching an upper limit based on the graph’s properties.

% % \vspace{-10pt}
% \begin{figure}
%     \centering
%     \includegraphics[width=\linewidth]{pic//exp/vp_impact}
%     % \vspace{-10pt}
%     \caption{Running time when varying the number of $k$DPs with $k$=$10$.}
%     \label{fig:vpcnt_impact}
% \end{figure}
% % \vspace{5pt}

% % 删除 end
    
\subsubsection{Ablation Study}\label{sec:experiment_opt}

We evaluate key components of \texttt{ShareDP} with $k$=$10$ on the 4 largest graphs in terms of running time (Tab.~\ref{tab:optimization_impact}, bold indicates the best, and underline indicates the second-best.).
(1) {\emph{Merged Split-Graph Representation.}}  
\texttt{ShareDP} outperforms \texttt{ShareDP-}, validating the effectiveness of our representation.
(2) {\emph{Merged Traversal.}}  
\texttt{ShareDP-} outperforms \texttt{maxflow} in 3 datasets, demonstrating the effectiveness of merged traversals. However, for the largest dataset, \texttt{maxflow} outperforms \texttt{ShareDP-}, highlighting the importance of well-designed components (e.g., representation of $G'$) beyond merged traversals.

% % % \vspace{-10pt}
% \begin{table}[!h]
%     \centering
%     \caption{Ablation Study. Average execution time (seconds) with $k$=$10$. }
%     % \resizebox{\linewidth}{!}
%     {
%         \input{tab/tab_optimization_impact}
%     }
%     \label{tab:optimization_impact}
% \end{table}
% % % \vspace{-10pt}

%!TEX root = main.tex

% \vspace{-10pt}
\section {Conclusions}
\label{sec:conclusion}
% \vspace{-5pt}
We study the problem of batch-$k$DP and propose the \texttt{ShareDP} algorithm. 
\texttt{ShareDP} shares computations across queries by 
    consolidating converted graphs into a shared structure and 
    sharing the traversals within this framework. 
Extensive experiments confirm the superiority of \texttt{ShareDP} over existing approaches.

% % \vspace{-10pt}
\begin{table}[!h]
    \centering
    \caption{Ablation Study. Average execution time (seconds) with $k$=$10$. }
    % \resizebox{\linewidth}{!}
    {
        \begin{tabular}{|c|c|c|c|c|}
\hline
Method & id & ar & uk & tw \\
\hline
\textit{ShareDP-} & \underline{ 6.27 } & \underline{ 19.16 } & \underline{ 26.77 } & 19.42 \\
\textit{ShareDP} & \textbf{ 2.54 } & \textbf{ 7.27 } & \textbf{ 6.16 } & \textbf{ 5.34 } \\
\textit{maxflow} & 15.53 & 35.24 & 101.89 & \underline{ 11.19 } \\
\hline
\end{tabular}
    }
    \label{tab:optimization_impact}
\end{table}
% % \vspace{-10pt}

% \vspace{-20pt}
% % \vspace{-25pt}
% \vspace{-20pt}

% \vspace{-10pt}
% \vspace{5pt}

\section*{Acknowledgement}
This work was supported by The National Key Research and Development Program of China under grant 2023YFB4502303. Lei Zou is the corresponding author of this work.

% \clearpage

% \balance

% \bibliographystyle{splncs04}
% \bibliography{main}

% \newpage
% \bibliographystyle{abbrv}
\printbibliography

\clearpage
\appendix
\section{Complete Algorithm of \texttt{ShareDP}} \label{sec:appendix}
\begin{algorithm}[H]
    \caption{ShareDP}
    \label{alg:shareDP}
    \KwIn{Set of $k$DP queries $Q$, original graph $G$, integer $k$}
    \KwOut{$k$ disjoint paths for each query $q \in Q$}
    
    Initialize \{$P$\}: Set $prehops$, $nexthops$, $isPinner$ to $\emptyset$, and initialize $isS$ and $isT$ according to the definition.\textit{\color{gray}// Initialize the merged split-graph. See Sec.~\ref{sec:approach_details}.} \\
    \ForEach{$i$ $\in$ $1, \ldots, k$ \textit{\color{blue}/* Find the $i$th disjoint paths */}} 
    {
        \textit{\color{blue}/* Initialization for bidirectional BFS */} \\
        Initialize $s$-$seen$, $s$-$queue$, $s$-$nextqueue$ for forward BFS. Backward BFS uses analogous structures (e.g., $t$-$queue$). \\
        Initialize $joint$, $pred$, and $succ$ \textit{\color{gray}// $joint$: meeting points of forward and backward searches; $pred$/$succ$: path predecessors/successors.} \\
        $undone = Q$ \textit{\color{gray}// Queries that have not yet found the $i$th path.} \\
        
        \ForEach{$q \in Q$} 
        {
            $s$-$seen_{q.s} \cup= \{q\}$; $s$-$queue_{q.s} \cup= \{q\}$ \textit{\color{gray}// Mark $q.s$ as visited and enqueue if for forward search. Initialization for backward search is similar.} \\
        }

        \textit{\color{blue}/* Bidirectional BFS */} \\
        \While{$undone \neq \emptyset$ and $s$-$queue$ and $t$-$queue$ are not empty} 
        {
            \textit{\color{gray}/* Forward search */} \\
            \ForEach{\{$v$, $B$\} $\in s$-$queue$ \textit{\color{gray} // $B$ is a set of $k$DPs that need to expand $v$}}
            {
                ForwardExpandFrontier($v$, $B$, $G$, \{$P$\}) \textit{\color{gray}// See Alg.~\ref{alg:expandFrontier}.}
                % $B = B \setminus undone$ \textit{\color{gray}// Skip queries that have already found the $i$th path.} \\
                % \ForEach{\{$u$, $B'$\} $\in$ GetOutNeighbors($v$, $B$, $G$, \{$P$\}) \textit{\color{gray}// See Alg.~\ref{alg:get_nbr}.}} 
                % {
                %         $D = B' \setminus s$-$seen_u$ \textit{\color{gray}// Exclude queries that have already visited $u$.} \\
                %         $s$-$seen_u \cup= D$; $pred_{u, v} \cup= D$ \textit{\color{gray}// Mark $u$ as visited for $D$ and record $v$ as the predecessor.} \\
                %         $meet = D \cap t$-$seen_u$ \textit{\color{gray}// Queries that forward search and backward search meet at $u$.} \\
                %         $joint_u \cup= meet$; $undone \setminus= meet$ \textit{\color{gray}// Mark queries as completed.} \\
                %         $s$-$nextqueue_u \cup= (D \setminus meet)$ \textit{\color{gray}// Add $u$ to the queue for remaining queries.} \\
                % }
            }
            Swap $s$-$queue$ and $s$-$nextqueue$ \\
            \textit{\color{gray}/* Backward search proceeds similarly to forward search. */}
        }

        \textit{\color{blue}/* Construct paths and update the merged split-graph */} \\
        Construct paths from $undone$, $joint$, $pred$, and $succ$. \\
        Update $prehops$, $nexthops$, and $isPinner$.\textit{\color{gray}// Adjust paths similar to flow augmentation in max-flow problem. For example, if trying to add edge $\langle u, v\rangle$ for queries $B$, first let $cancel = prehops_{u, v} \cap B$ (edge $\langle v, u\rangle$), then remove $cancel$ from $prehops_{u, v}$ and add $(B \setminus cancel)$ to $prehops_{v, u}$. Similarly, update $nexthops$.} \\
    }
    
    Extract $k$ disjoint paths for each $q \in Q$ from $isS$ and $nexthops$ (or $isT$ and $prehops$) and return the results. \\
\end{algorithm}

\end{document}